\documentclass[floatfix,lengthcheck,showpacs,amssymb,amsmath,amsfonts,twocolumn,nofootinbib,longbibliography]{revtex4-1}

\usepackage{lineno}
\usepackage{color}
\usepackage{graphicx}
\usepackage{float}
\usepackage{subfigure}
\usepackage{amsmath}
\usepackage{multirow}
\usepackage{tikz}
\usetikzlibrary{arrows,shapes,trees,decorations.pathreplacing}
\usepackage{import}
\usepackage{svn-multi}
\usepackage{hyperref}
\usepackage{gensymb}
\hypersetup{
    colorlinks=true,
    linkcolor=blue,
    filecolor=magenta,      
    urlcolor=cyan,
}
\DeclareUnicodeCharacter{2009}{\,}

\newcommand{\msun}{\ensuremath{\mathrm{M}_{\odot}}}
\newcommand{\mjup}{\ensuremath{\mathrm{M}_{J}}}

\usepackage{acro}
\DeclareAcronym{GR}{
	short = GR,
	long  = general relativity
	}
\DeclareAcronym{SNR}{
	short = SNR,
	long  = signal-to-noise ratio
	}
	
\DeclareAcronym{BBH}{
	short = BBH,
	long  = binary black hole
	}
	
\DeclareAcronym{PBH}{
short = PBH,
long  = primordial black hole
}

\begin{document}

\title{Search for gravitational waves from high-mass-ratio compact-binary mergers of stellar mass and sub-solar mass black holes}

\author{Alexander H. Nitz}
\email{alex.nitz@aei.mpg.de}
\author{Yi-Fan Wang}
\affiliation{Max-Planck-Institut f{\"u}r Gravitationsphysik (Albert-Einstein-Institut), D-30167 Hannover, Germany}
\affiliation{Leibniz Universit{\"a}t Hannover, D-30167 Hannover, Germany}

\begin{abstract} 
We present the first search for gravitational waves from the coalescence of stellar mass and sub-solar mass black holes with masses between 20 - 100 \msun{} and 0.01 - 1 \msun{}($10 - 10^3$ \mjup), respectively. The observation of a single sub-solar mass black hole would establish the existence of primordial black holes and a possible component of dark matter. We search the $\sim 164$ days of public LIGO data from 2015-2017 when LIGO-Hanford and LIGO-Livingston were simultaneously observing. 
We find no significant candidate gravitational-wave signals. Using this non-detection, we place a $90\%$ upper limit on the rate of $30-0.01~\msun$ and $30-0.1~\msun$ mergers at $<1.2\times10^{6}$ and $<1.6\times10^{4} ~\mathrm{Gpc}^{-3} \mathrm{yr}^{-1}$, respectively.
If we consider binary formation through direct gravitational-wave braking, this kind of merger would be exceedingly rare if only the lighter black hole were primordial in origin ($<10^{-4}~\mathrm{Gpc}^{-3}\mathrm{yr}^{-1}$). 
If both black holes are primordial in origin, we constrain the contribution of $1 (0.1)~\msun$ black holes to dark matter to $< 0.3 (3)\%$. 
\end{abstract}

\date{\today}
\maketitle

\section{Introduction}
The first gravitational wave observation from the merger of black holes was detected on September 14th, 2015~\cite{abbott:2016blz}. Over a dozen \ac{BBH} mergers~\cite{Nitz:2018imz,Nitz:2019hdf,Nitz:2020naa,Venumadhav:2019tad,Venumadhav:2019lyq,Zackay:2019btq,LIGOScientific:2018mvr} have since been reported along with two binary neutron star mergers~\cite{TheLIGOScientific:2017qsa,Abbott:2020uma} by Advanced LIGO~\cite{TheLIGOScientific:2014jea} and Virgo~\cite{TheVirgo:2014hva}. Recently, two compact binary coalescences with unequal masses have been reported~\cite{LIGOScientific:2020stg,GW190814}; the mass ratios are $\sim 3$ and $\sim 9$, respectively.

The nature of dark matter remains a mystery given null results from direct searches using particle experiments (see e.g., Refs. \cite{PhysRevLett.119.181302,PhysRevLett.120.061802} and recent notable exception in Ref.~\cite{Aprile:2020tmw}).
The observation of \ac{BBH} mergers has sparked renewed interest in primordial black holes (PBHs) as a contributor to dark matter~\cite{Hawking:1971ei,Carr:1974nx,PBH1993-Silk,PBH1996,Bird:2016dcv,Clesse:2016vqa,Sasaki:2016jop,Chen:2018czv,DeLuca:2020qqa}. 
However, the merger of stellar-mass PBHs may be difficult to separate from standard stellar formation channels. Black holes may form through standard stellar evolution between $~2-50 \msun$~\cite{2012ApJ...749...91F,2014ApJ...785...28K,Ertl_2016,Woosley:2016hmi,Marchant:2018kun,Farmer:2019jed}. Furthermore, gravitational-wave observation alone is not always able to determine if a component of a binary is either a neutron star or black hole~\cite{Yang:2017gfb,LIGOScientific:2019eut}. Although, the observation of coincident gamma-ray bursts or kilonovae, such as in the case of GW170817~\cite{TheLIGOScientific:2017qsa,Monitor:2017mdv,GBM:2017lvd,Soares-Santos:2017lru,Coulter:2017wya}, can confirm the presence of nuclear matter. In contrast, there is no known model which can produce sub-solar mass black holes by conventional formation mechanisms; the observation of a single sub-solar mass black hole would provide strong evidence for PBHs.

There are a variety of constraints for the contributing fraction of PBHs to dark matter (see Refs.~\cite{Carr:2020xqk,Carr:2020gox} and references therein). Gravitational-wave astronomy provides a unique window; notably, a direct search for sub-solar mass black holes constrained the mass range $0.2-2$~\msun{} for near equal-mass sources~\cite{Authors:2019qbw, Abbott:2018oah} and the non-detection of a gravitational-wave astrophysical background by LIGO and Virgo has constrained PBHs with $0.01-100$~\msun{}~\cite{Wang:2016ana}. Recently, tight constraints from the NANOGrav pulsar timing array~\cite{Arzoumanian:2017puf} are given by Ref.~\cite{Chen:2019xse} for $0.001-1$~\msun{} black holes based on the non-detection of gravitational waves induced by scalar perturbations during the expected PBH formation epoch.

\begin{figure}[tb!]
    \centering
    \includegraphics[width=\columnwidth]{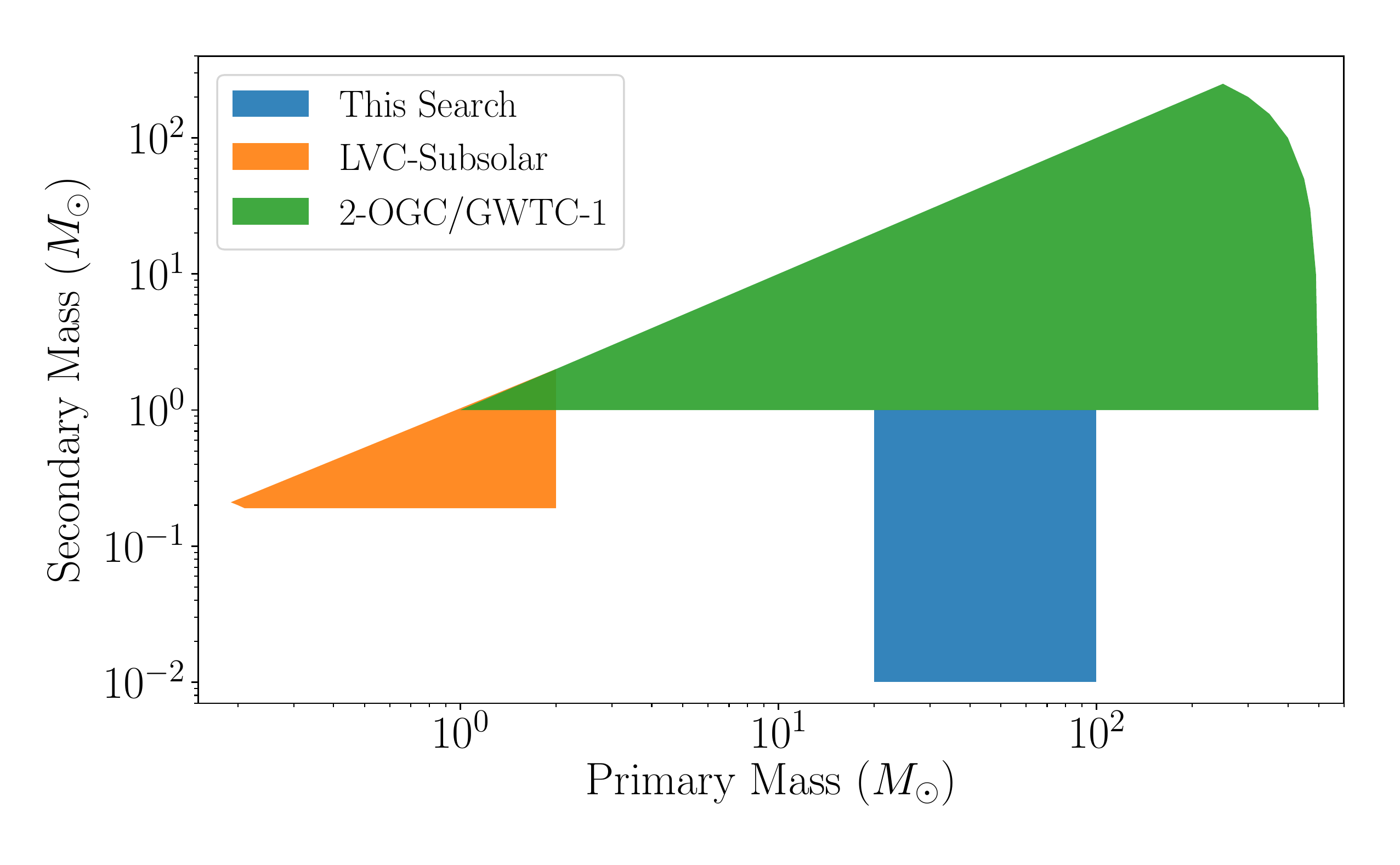}
    \caption{The primary and secondary masses of the sources searched by our analysis (blue), 2-OGC/GWTC-1 (green)~\cite{Nitz:2019hdf,LIGOScientific:2018mvr}, and the sub-solar mass LVC search (orange)~\cite{Authors:2019qbw}. }
    \label{fig:searches}
\end{figure}

So far, all observations of gravitational waves from BBH mergers were identified by searches targeting stellar-mass sources, i.e. those with component masses $1-\mathcal{O}(100)~\msun$. Targeted searches for sub-solar mass binaries with component masses between $0.2-2~\msun$ have null results~\cite{Authors:2019qbw, Abbott:2018oah}. We report a search for sub-solar mass black holes in an unexplored region of parameter space: the merger of $0.01-1~\msun$ sub-solar mass black holes with $20-100~\msun$ stellar-mass black holes. We summarize the region we search in comparison to past analyses in Fig.~\ref{fig:searches}. We find no statistically significant candidates and place the first constraints from gravitational-wave observation on the merger rate of these sources.

\section{Search}

\begin{figure}[tb!]
    \centering
    \includegraphics[width=\columnwidth]{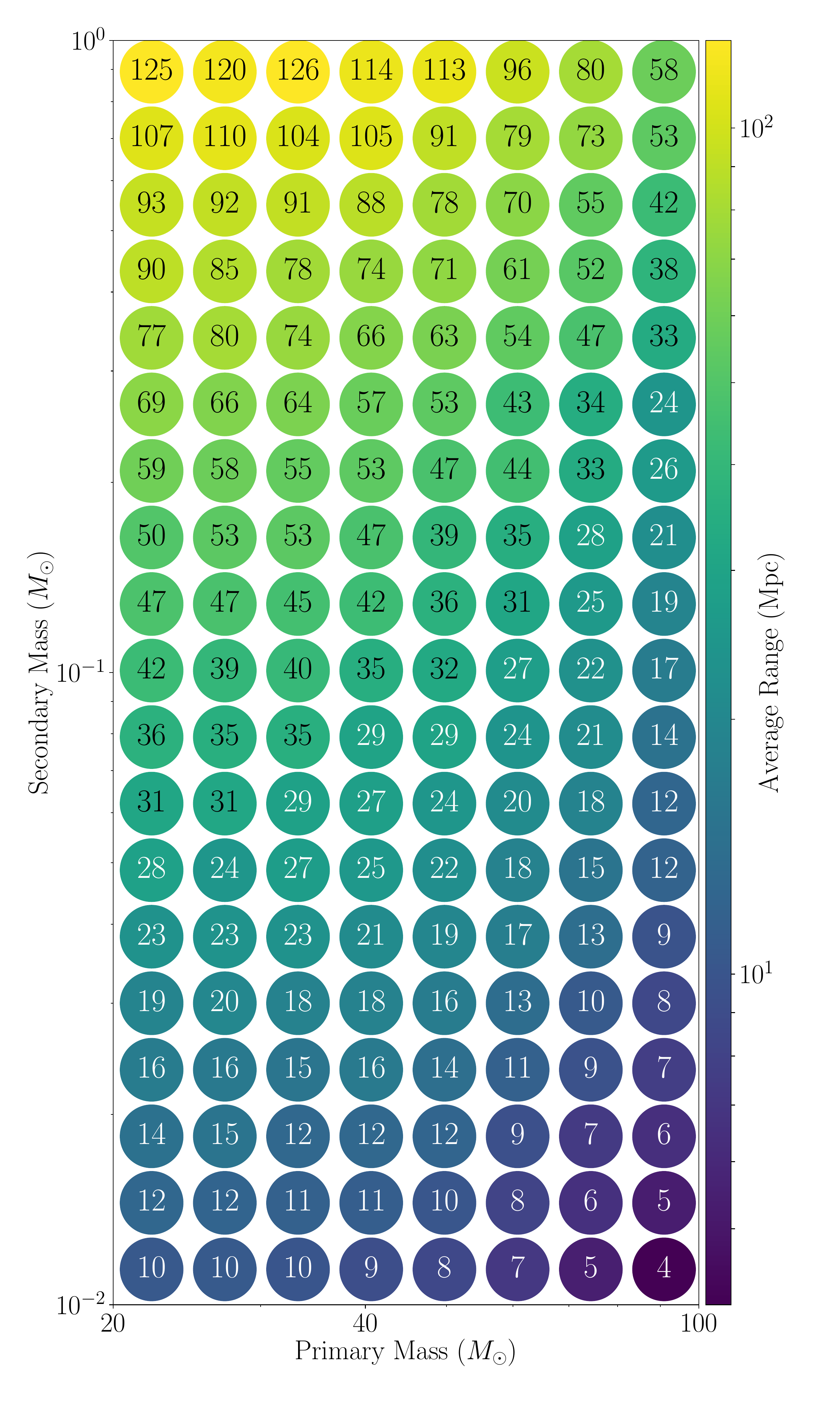}
    \caption{The distance our search can detect sources at a false alarm rate of 1 per 100 years as a function of the primary and secondary masses, averaged over the possible sky locations and orientations of an isotropic source population, and averaged over the observation period. The horizon distance, the maximum distance a source could be found, is a factor of $\sim2.26$ larger than the average range shown here.}
    \label{fig:sense}
\end{figure}

We analyze the public LIGO data from 2015-2017, which contains $\sim164$ days of joint LIGO-Hanford and LIGO-Livingston observation time~\cite{Vallisneri:2014vxa,Abbott:2019ebz}. Virgo was observing for the final month of this period, but had limited range in comparison to the LIGO instruments. We use the open-source PyCBC-based search pipeline~\cite{Usman:2015kfa,pycbc-github} configured similarly to the analysis of Ref.~\cite{Nitz:2019hdf} to analyze the LIGO data, identify potential candidates, apply tests of each candidate's signal consistency~\cite{Allen:2004gu,Nitz:2017lco}, rank each candidate, and finally assess each candidate's statistic significance~\cite{Nitz:2017svb,Davies:2020,Simone:2020}. The statistical significance of any candidate is assessed by comparing to the empirically estimated rate of false alarms. This rate is estimated by creating numerous fictitious analyses analyzed in an identical manner to the search, but where time shifts between the data of the two LIGO observatories are applied. The time shift of each background analysis is greater than the light-travel-time between the two LIGO observatories, which ensures astrophysical signals are not in coincidence. The average sensitive distance of our analysis at a false alarm rate of 1 per 100 years is shown in Fig.\ref{fig:sense}.

Matched filtering is used to extract the signal-to-noise from data for a given template waveform~\cite{Usman:2015kfa,gstlal-methods}. Each template corresponds to the gravitational-wave signal of a single source type. To search for sources with varied component masses, a discrete bank of template waveforms is required. We use a brute-force stochastic method~\citep{Harry:2009ea} to find the nearly 9 million templates required by our analysis. To reduce computational cost, we search at most the final 60 seconds of each gravitational waveform. For the lightest sources, this implies we analyze the data starting at a higher gravitational-wave frequency than for the heaviest sources, where we analyze the data starting from 20 Hz.

\begin{figure*}[tb!]
    \centering
    \includegraphics[width=2\columnwidth]{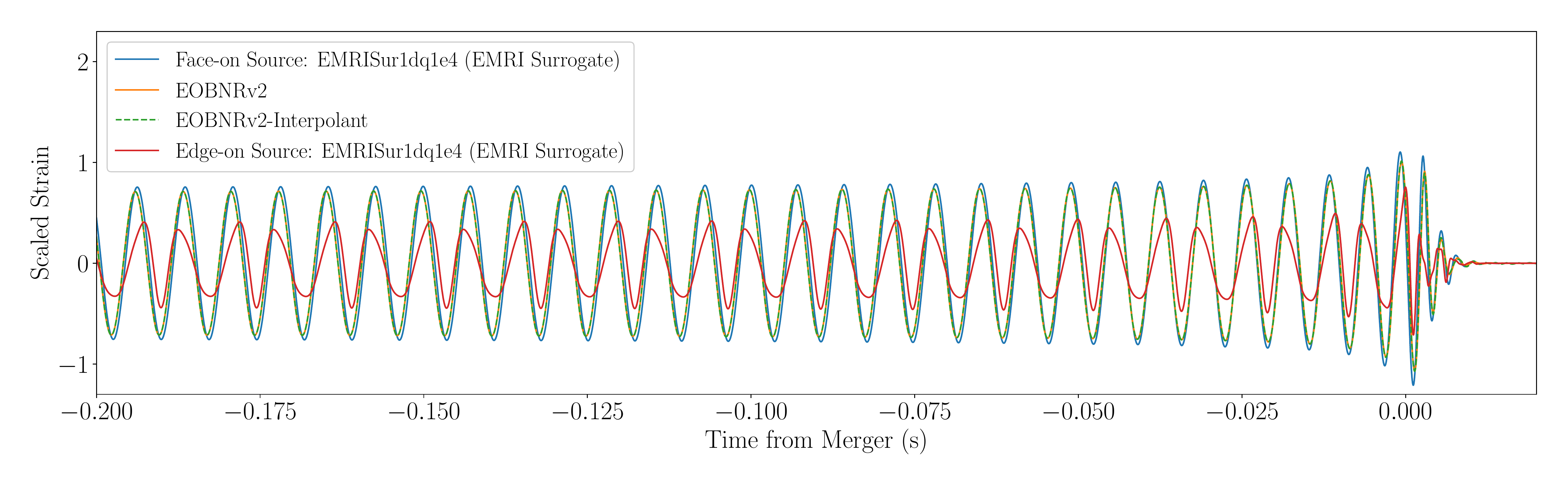}
    \caption{Comparison of the gravitational waveform for a 30~\msun - 0.01~\msun{} merger. The EOBNRv2 interpolant model used by our search is consistent with the EMRISur1dq1e4 surrogate model when the inclination of the source's orbital plane is close to face-on/off. For sources with highly inclined orbital planes, higher order modes becomes increasingly important.}
    \label{fig:waveform}
\end{figure*}

To model the gravitational-wave signal, we use EOBNRv2, a model based on an effective one-body Hamiltonian approximation of general relativity in combination with a fitted merger and ringdown~\cite{pan:2009wj}. Several phenomenological models exist for BBH mergers, however, they do not generalize to the high mass ratios we consider~\cite{Pratten:2020ceb,Khan:2015jqa}. We assume our sources' orbits have negligible eccentricity by the time of observation and that the component black holes are non-spinning. This choice is consistent with the prediction that PBHs have negligible component spin~\cite{Chiba:2017rvs,DeLuca:2019buf,DeLuca:2020bjf,Mirbabayi:2019uph}. Due to the degeneracy between mass ratio and spin~\cite{Baird:2012cu}, we expect our search to be able to recover moderately spinning sources where $\chi_{1,2}\lesssim 0.1$~\cite{Brown:2012gs,Nitz:2013mxa}, which is well beyond the larger predictions at the percent level~\cite{Postnov:2019pkd}. EOBNRv2 is too slow for use by our search directly. Instead, we use a straightforward interpolant based on $\sim10^4$ pre-generated EOBNRv2 waveforms with different mass ratios which can be rapidly scaled to any point in parameter space. We crosscheck our model against the recent extreme mass ratio inspiral (EMRI) surrogate EMRISur1dq1e4~\cite{Rifat:2019ltp} and find that our interpolant, the base EOBNRv2 model, and the dominant-mode of EMRISur1dq1e4 have less than $<1\%$ mismatch at all locations in our template bank, i.e any of these models would recover $>99\%$ of the SNR of a signal matching one of the other models. A visual comparison between a representative example of these models is in Fig.~\ref{fig:waveform}. 

 Gravitational-waves are expressed in terms of spin-weighted spherical harmonic decomposed modes. Methods exist for incorporating higher order modes into searches at increased computation cost~\cite{Harry:2017weg}. EOBNRv2 provides only the dominant mode of the gravitational waveform, and accurate models with higher order modes exist only for lower mass ratio sources~\cite{Pan:2011gk,London:2017bcn,Cotesta:2020qhw}, or short duration signals~\cite{Rifat:2019ltp}. We compare our templates against these models to estimate the potential loss in search sensitivity. Neglecting higher order modes in our search imposes an source-orientation averaged loss in \ac{SNR} of $\sim5, 10, 15, 22\%$ for sources with a 20, 40, 60, and 100 $\msun$ primary mass, respectively. The most significant loss in sensitivity is for sources with near edge-on inclination of their orbital plane with respect to an observer, whereas higher order modes become negligible for sources with near face-on inclination. The search sensitivity and upper limits quoted in this paper account for the detection rate reduction.

\section{Observational Results}

The most significant candidate from our search was observed at a false alarm rate of 3 per year~\footnote{Additional details about the most significant candidates can be found at \url{https://github.com/gwastro/stellar-pbh-search}.}, and if it were astrophysical, would be consistent with the merger of a $23~\msun$ primary black hole with a $0.012~\msun$ secondary. Considering the time searched, our results are consistent with a null observation.

\begin{figure}[tb!]
    \centering
    \includegraphics[width=\columnwidth]{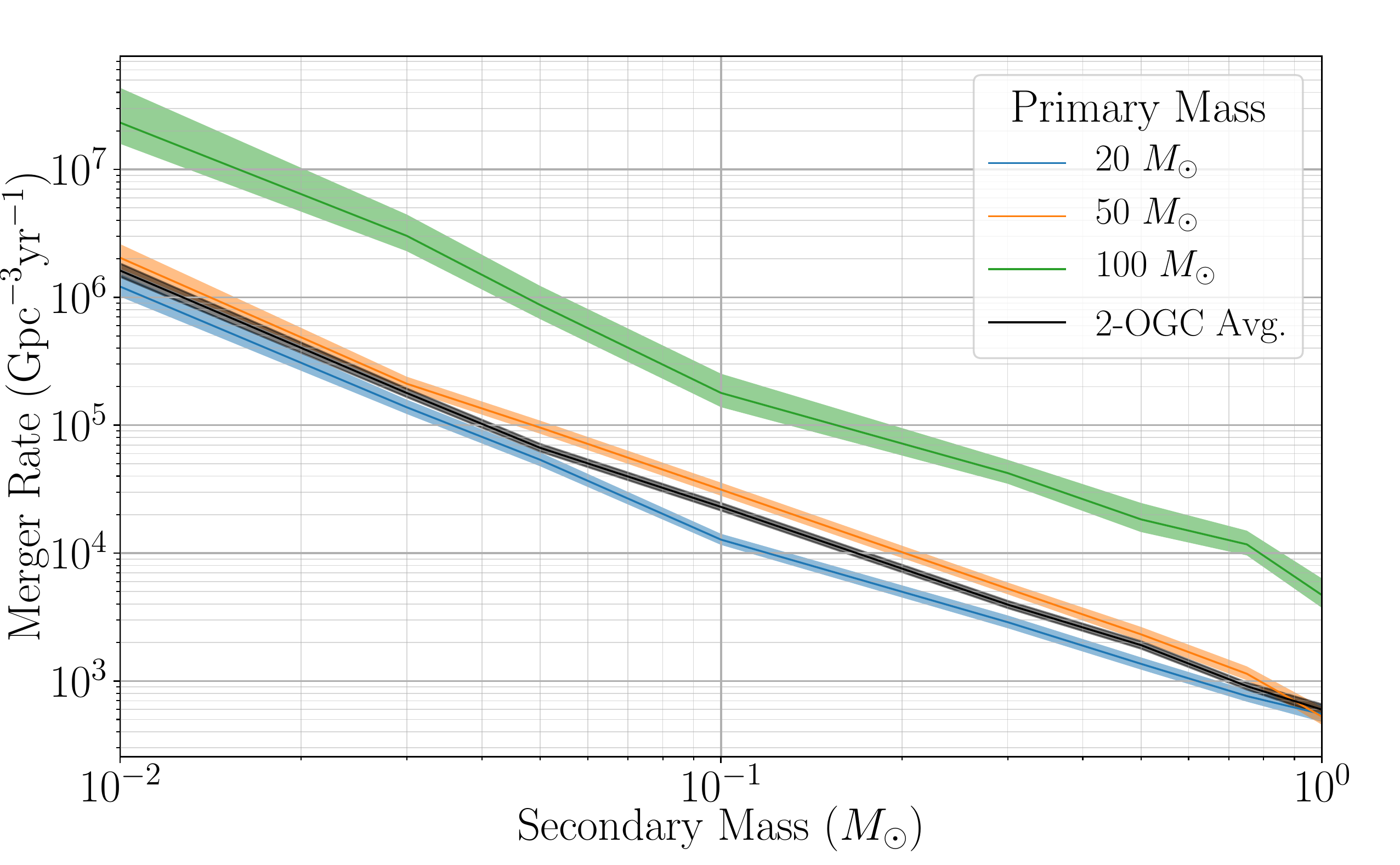}
    \caption{The $90\%$ upper limit on the rate of mergers as a function of the mass of the secondary black hole, for a range of primary masses (various colors), and the average assuming a primary mass population consistent with observed BBH mergers from the 2-OGC catalog (black)~\cite{Nitz:2019hdf}. The one sigma monte-carlo statistical uncertainty is shown with shading.}
    \label{fig:rate}
\end{figure}

We place upper limits at 90\% confidence on the rate of mergers throughout the searched space using the loudest event method~\cite{Biswas:2007ni}. The upper limit $R_{90}$ is given as,
\begin{equation}
    R_{90} = \frac{2.3}{VT}
\end{equation}
where V is the sensitive volume of our analysis at the false alarm rate of the most significant candidate, and T is the total time searched. We simulate a population of sources distributed isotropically in the sky and binary orientation, and uniform in volume, to measure the sensitive volume of our analysis as a function of the primary and secondary masses. Figure~\ref{fig:rate} shows the upper limit on the merger rate as a function of the secondary mass. Assuming a distribution of primary masses consistent with the black holes observed by LIGO and Virgo, we find that the rate of $0.01~\msun$ solar mass mergers is $<1.7\times10^6~\mathrm{Gpc}^{-3} \mathrm{yr}^{-1}$ at 90\% confidence.

\section{Implications for Primordial Black Holes}

Whereas stellar-mass black holes can be either the product of stellar evolution or PBHs, sub-solar mass black holes only form in the primordial Universe given conventional stellar evolution \cite{Carr:2020xqk,Carr:2020gox}. There are two possible origins for the high-mass-ratio binaries we considered; the first that only the secondary, lighter black hole is primordial and forms a binary with a stellar-origin black hole in the late Universe, and a second scenario where both black holes are primordial and formed a binary in the early Universe.

For the first scenario, binaries form in the galactic field through dynamical capture due to gravitational-wave bremsstrahlung. For PBHs with mass $m_1$, stellar-origin black holes with mass $m_2$ and relative velocity $v$, the cross section for binary formation is given by Ref.~\cite{Mouri:2002mc} as
\begin{equation}
    \sigma = 2\pi \left(\frac{85\pi}{6\sqrt{2}}\right)^{2/7}\frac{G^2(m_1+m_2)^{10/7}(m_1m_2)^{2/7}}{c^{10/7}v^{18/7}}
\end{equation}
where $G$ and $c$ are the gravitational constant and speed of light, respectively. As shown by Ref.~\cite{Bird:2016dcv}, binaries are expected to quickly merge after formation and disruption by other PBHs can be neglected.

To estimate the PBH distribution, we use a recent cosmological galaxy formation simulation \textit{IllustrisTNG}~\cite{illustris}.
In the $\mathrm{redshift}=0$ snapshot of the TNG-100 high resolution simulation, there are $\sim 10^5$ dark matter main subhalos with non-zero star formation within a $\sim 100$ Mpc size cube.
For each main subhalo, we assume the dark matter number density follows the Navarro-Frenk-White (NFW) profile $\rho_\mathrm{NFW}$ \cite{NFW}, and that PBHs constitute a fraction of dark matter with mass fraction $f_\mathrm{PBH}$.

Estimating the abundance of stellar-origin black holes is a challenge due to the lack of observation. Nevertheless, the synthesis population study of Ref.~\cite{Olejak:2019pln} shows that $\sim0.006\%$ of the total galactic halo mass including dark matter is in the form of black holes. As an approximation, we take this value as the universal fraction over dark matter main subhalos to infer the mass density $\rho_\mathrm{BH}$ of stellar-origin black holes. 

The rate density of dynamical captures between primordial and stellar-origin black holes is finally
\begin{equation}
   R(m_1,m_2) =  \sum_\mathrm{Halos}  
        \int_0^{\sqrt[3]{2}R_\mathrm{halfmass}} \frac{\rho_\mathrm{BH}}{m_1}
        \frac{f_\mathrm{PBH}\rho_\mathrm{NFW}(r)}{m_2} 
        \sigma v  d^3 r . 
        \label{eq:ratescenario1}
\end{equation}
Assuming a uniform spatial distribution of stellar-origin black holes, the radius $r$ is integrated from the main subhalo center to $\sqrt[3]{2}$ times of the radius which contains half of the stellar mass, $R_\mathrm{halfmass}$. The relative velocity $v$ is approximated by the stellar dispersion velocity, provided by \textit{IllustrisTNG}. We find that even for $f_\mathrm{PBH}=100\%$, this formation channel implies a merger rate $<10^{-4}$ Gpc$^{-3}$ yr$^{-1}$ for $37~M_\odot-0.01~M_\odot$ binaries. 

The estimation of $\rho_\mathrm{BH}$ is a source of uncertainty, however, other variables in Eq.~\ref{eq:ratescenario1} such as dark matter halo abundance are not expected to change by orders of magnitude since they are extracted from the robust simulation. Given that the resultant rate is $\sim 10$ orders of magnitude lower than that shown in  Fig.~\ref{fig:rate}, our conclusion that this scenario is unlikely is robust.

\begin{figure}[tb!]
    \centering
    \includegraphics[width=\columnwidth]{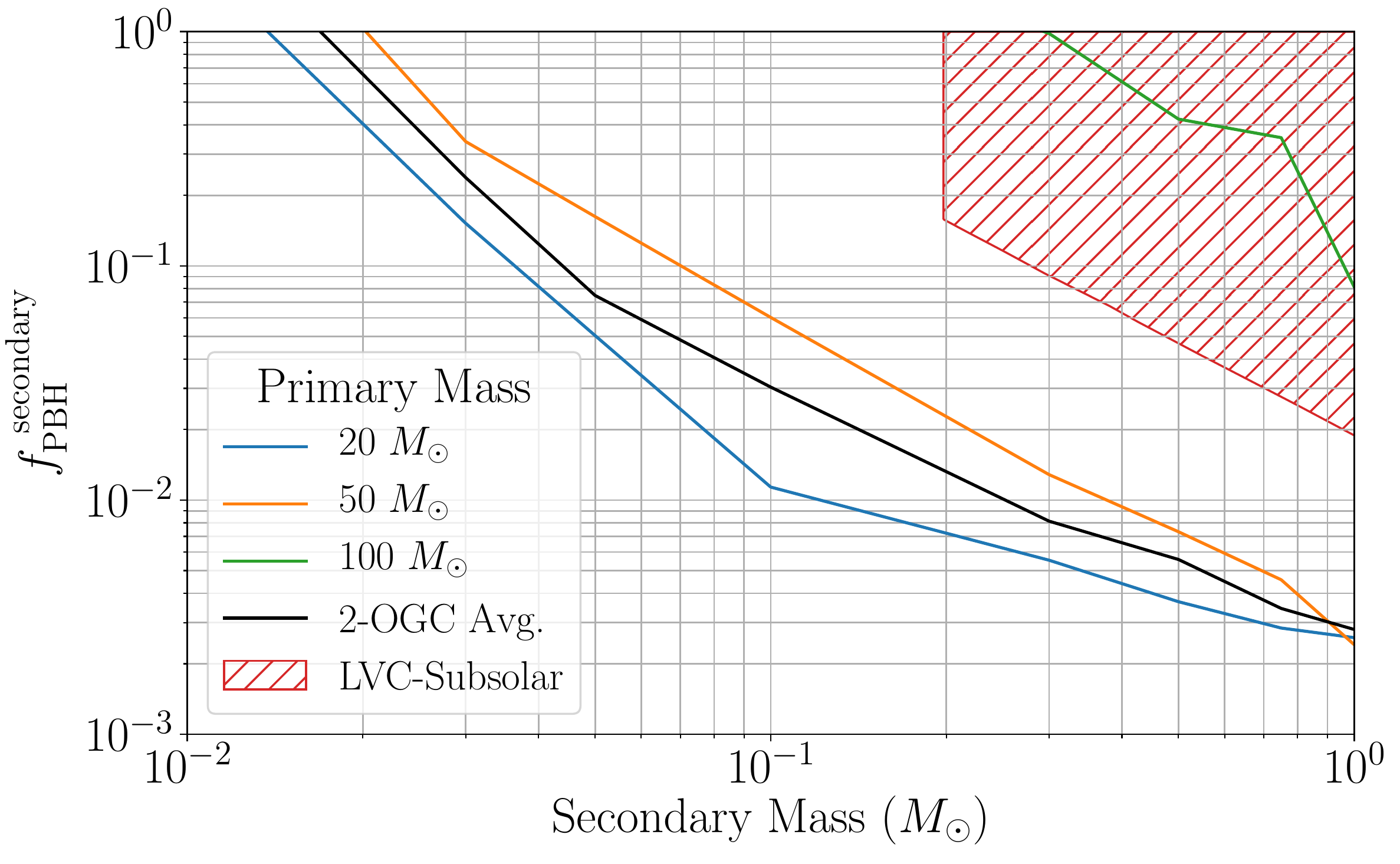}
    \caption{Upper limits on $f_\mathrm{PBH}^\mathrm{secondary}$ for the secondary, sub-solar mass black hole assuming both primary and secondary black holes have primordial origin, where we choose the primary mass to be $20/50/100 \msun$ (blue/orange/green) or the average mass of the 2-OGC catalog ($\sim 37$ \msun) (black). The constraints from the LVC direct search for equal-mass PBHs~\cite{Authors:2019qbw} are plotted for comparison.
    }
    \label{fig:pbh}
\end{figure}


On the other hand, if both primary and secondary black holes are PBHs, a nearest pair may form a binary if decoupled from the Universe's expansion. Refs.~\cite{Chen:2018czv,Ali-Haimoud:2017rtz} consider a uniform spatial distribution when PBHs form and give the merger rate for a binary with mass $m_1$ and $m_2$ as
\begin{eqnarray}\label{eq:pbhrate}
&R(m_1,m_2)& = 3.3\cdot10^6 \cdot f_\mathrm{PBH}^2(0.7f_\mathrm{PBH}^2 + \sigma_\mathrm{eq}^2)^{-\frac{21}{74}}(m_1m_2)^{\frac{3}{37}}\nonumber \\
&\times(m_1+m_2)^{\frac{36}{37}}& \mathrm{min}\left(\frac{P(m_1)}{m_1},\frac{P(m_2)}{m_2}\right
)\left(\frac{P(m_1)}{m_1}+\frac{P(m_2)}{m_2}\right)
\end{eqnarray}
where mass $m$ and merger rate $R$ are in units of \msun{} and Gpc$^{-3}$ yr$^{-1}$, respectively. $P(m)$ is the normalized PBH mass distribution. 
The parameter $\sigma_{eq}$ accounts for the variance of density perturbation from other dark matter aside from PBHs at the matter radiation equality epoch and is suggested to be $0.005$ by Ref.~\cite{Ali-Haimoud:2017rtz}. 

The possibility that the currently observed stellar-mass \ac{BBH} mergers were caused by PBHs is a topic of investigation~\cite{DeLuca:2020qqa,Bird:2016dcv,Clesse:2016vqa,Sasaki:2016jop,Chen:2018czv,Chen:2019irf}.
In the most optimistic case, where the majority of LIGO and Virgo observed black holes are PBHs, $f_\mathrm{PBH}^\mathrm{primary}=3 \times 10^{-3}$ by Ref.~\cite{Chen:2019irf}.
With this fixed fraction for the primary mass, we constrain the contribution of the secondary, sub-solar mass black hole to dark matter. We assume a two-valued mass distribution, i.e., $P(m_1) + P(m_2) = 100\%$. Thus the fraction in dark matter for the primary and secondary black hole is $f_\mathrm{PBH}^\mathrm{primary}=P(m_1)f_\mathrm{PBH}$ and $f_\mathrm{PBH}^\mathrm{secondary}=P(m_2)f_\mathrm{PBH}$.

The upper limit for $f_\mathrm{PBH}^\mathrm{secondary}$ for a fixed fiducial primary mass $m_1=20/50/100$ \msun{} and the average mass from the 2-OGC catalog ($\sim$37 \msun)~\cite{Nitz:2019hdf} are shown in Fig.~\ref{fig:pbh}. For the 2-OGC average case, we find that 1(0.1) \msun{} PBHs cannot exceed $0.3(3)\%$ of the total dark matter. In contrast, if we assume none of the LIGO/Virgo \ac{BBH} detections are PBHs, our results cannot constrain $f_{\mathrm{PBH}}$.

Our constraints can be directly compared with the targeted search for near equal-mass sub-solar black holes~\cite{Authors:2019qbw, Abbott:2018oah} which used the same formation scenario as described by Eq.~\ref{eq:pbhrate}. Our results expand the range probed by direct merger observation down to to $0.01 M_\odot$. The constraint on the abundance of PBHs is improved by an order of magnitude as we consider sources with higher total mass that emit stronger gravitational waves.

A significant source of uncertainty in this model derives from the fraction of binaries which survive to merger. Under active investigation is what fraction of PBH binaries would be disrupted after initial formation. If a significant fraction are disrupted, the event rate would be lowered and our constraints loosened. Ref.~\cite{Ali-Haimoud:2017rtz} has shown analytically that the disruption can be neglected, however recently Ref.~\cite{Jedamzik:2020omx} argues for a higher disruption rate. Further, if PBHs exhibit substantial clustering at formation, the event rate may be boosted and our constraints would be tighter~\cite{Bringmann:2018mxj,Ballesteros:2018swv}.  As we consider the same model, both our results and LVC limits shown in Fig.~\ref{fig:pbh} don't consider binary disruption and assume a uniform spatial distribution when PBHs form.

Stringent constraints for sub-solar mass PBHs from pulsar timing arrays~\cite{Chen:2019xse} have almost excluded the $0.001-1$~\msun{} mass region, overlapping with our $0.01-1$~\msun{} region. However, the scalar induced gravitational waves considered in Ref.~\cite{Chen:2019xse} apply to \textit{Gaussian} scalar curvature perturbation in the early Universe. Refs.~\cite{Nakama:2016gzw,Cai:2019elf} have shown that non-Gaussianity can suppress the scalar induced gravitational waves by orders of magnitude depending on the detailed form of the perturbations. Positive results from a direct search for sub-solar mass compact objects would imply large local non-Gaussianity of primordial perturbation to alleviate the tension. Null results may also have implications for early Universe non-Gaussianity, but a detailed analysis is beyond the scope of this work.

\section{Conclusions}
We conduct a search for a previously unconsidered source of gravitational waves: the binary coalescence of high-mass-ratio sources, where the primary mass is 20-100~\msun{} and the secondary mass is 0.01-1~\msun{}. We find no promising candidates, and thus place improved upper limits on the merger rate and the abundance of PBHs.

The merging of a PBH with a black hole formed through stellar evolution is extremely unlikely under the scenario of direct capture through gravitational-wave braking. A significantly more efficient binary formation mechanism would be required for this scenario to make a significant contribution. On the other hand, assuming both black holes are primordial in origin places constraints on the abundance of PBHs.

Advanced LIGO and Virgo are continually being upgraded~\cite{livingreviewligo}, and the third generation of gravitational-wave detectors can further improve the horizon distance by an order of magnitude~\cite{et,cewhitepaper}. At that point, it will be possible to probe the redshift evolution of stellar-mass binaries to distinguish primordial and stellar-origin black hole distributions~\cite{Chen:2019irf}. From our results, we expect the constraint on sub-solar mass PBH abundance to be $~10^{3-4}$ times tighter than the current search, assuming a null result. Combining the results of ground-based detectors, pulsar timing, and possibly space-based detectors in the future, can together probe the existence of PBHs and may investigate the structure of primordial perturbations in the early Universe~\cite{Green:2020jor}.

\acknowledgments

 We acknowledge the Max Planck Gesellschaft and the Atlas cluster computing team at AEI Hannover for support.  This research has made use of data, software and/or web tools obtained from the Gravitational Wave Open Science Center (https://www.gw-openscience.org), a service of LIGO Laboratory, the LIGO Scientific Collaboration and the Virgo Collaboration. LIGO is funded by the U.S. National Science Foundation. Virgo is funded by the French Centre National de Recherche Scientifique (CNRS), the Italian Istituto Nazionale della Fisica Nucleare (INFN) and the Dutch Nikhef, with contributions by Polish and Hungarian institutes.
 
The top candidates from our analysis along with the configuration files necessary to reproduce the search are available at \url{https://github.com/gwastro/stellar-pbh-search}.

\bibliography{references,emri-pbh,sgwb_pbh,parity}

\end{document}